\documentclass[11pt]{article}
\usepackage{amsmath,amsfonts,latexsym,graphicx,amssymb}
\date{}
\def\la{\langle\,}
\def\r{\,\rangle}

\def\con{{}_{\_\rule{-1pt}{0pt}\_}
\rule{-2pt}{0pt}\raise1.5pt\hbox{$\mid$}\hspace{2pt}}

\title{\bf Phase-space approach to Berry's phases}
\author{Dariusz Chru\'sci\'nski\\
 Institute of Physics, Nicolaus Copernicus University\\
 ul. Grudzi\c{a}dzka 5/7, 87-100 Toru\'n, Poland}

\begin{document}

\maketitle

\begin{abstract}

We propose a new formula for the adiabatic Berry phase which is
based on phase-space formulation of quantum mechanics. This
approach sheds a new light into the correspondence between
classical and quantum adiabatic phases
--- both phases are related with the averaging procedure:
Hannay's angle with averaging over the classical torus and Berry's
phase with averaging over the entire classical phase space with
respect to the corresponding Wigner function. Generalizations to
the non-abelian Wilczek--Zee case and mixed states are also
included.

\end{abstract}

\noindent Geometric Berry phase \cite{Berry} and its classical
analog, so called Hannay angle, \cite{Hannay} (see also
\cite{Berry-clas}) have found numerous applications in various
branches of physics (see e.g. \cite{Wilczek} and \cite{book}).
Recently, it turned out that adiabatic Berry phase plays important
role in quantum computation algorithms as a model of a quantum
gate in a quantum computer (see e.g. \cite{Zanardi,Fahri}). In
this paper we present a new formula for the Berry  phase which is
based on the phase space formulation of quantum mechanics. This
approach sheds a new light into the correspondence between
classical and quantum adiabatic phases.

Both Berry's phase and Hannay's angles have been introduced in the
context of adiabatic evolution in quantum and classical mechanics,
respectively. Let us consider for simplicity a classical system
with one degree of freedom and let the corresponding phase space
be parameterized by canonical coordinates $(q,p)$. Suppose, that a
Hamiltonian $H(q,p;{\bf X})$ depends on a set of some external
parameters $\bf X$ from the parameter space $\cal M$ and that $\bf
X$ are changed adiabatically along a circuit $C$ and come back to
their initial values, i.e. ${\bf X}(T) = {\bf X}(0)$ for some
$T>0$. Now, the classical adiabatic theorem \cite{Arnold} states
that the system will evolve on the torus defined by the constant
value of the action variable $I$ and the angle variable varies
according to
\begin{equation}\label{}
  \theta(T) = \int_0^T \omega(I;{\bf X}(t))dt +
  \Delta\theta(I;C)\ ,
\end{equation}
where the frequency $\omega(I,X) = \partial H(I;X)/\partial I$ and
the additional shift --- Hannay angle $\Delta\theta$ --- is given
by the following integral over an arbitrary two-dimensional region
$\Sigma$ in $\cal M$ such that $C=\partial \Sigma$
\begin{equation}\label{}
 \Delta\theta(I;C) = - \frac{\partial}{\partial I}\,
 \int\!\int_{\partial\Sigma=C} \, F^{\rm c}(I;X) \ ,
\end{equation}
where $F^{\rm c}(I;{\bf X})$ denotes the following two-form on
$\cal M$:
\begin{equation}\label{}
F^{\rm c}(I;{\bf X}) =  \la d_{\bf X}\, p(I;{\bf X}) \wedge d_{\bf
X}\, q(I;{\bf X})\r \ ,
\end{equation}
and $\la f(I)\r$ denotes the average of $f(I,\theta)$ over the
torus $I$.

 Now, let us consider the quantized
system defined by $\hat{H}({\bf X})$. Clearly, the quantization $H
\rightarrow \hat{H}$ is not unique and depends on the ordering of
$\hat{q}$ and $\hat{p}$. In what follows we assume the
Wigner--Weyl (or symmetric) ordering, i.e. for example  $qp
\rightarrow (\hat{q}\hat{p} + \hat{p} \hat{q})/2$. The quantum
adiabatic theorem \cite{Messiah} states that a system originally
in an eigenstate $|n;{\bf X}(0)\r$ will remain in the same
eigenstate $|n;{\bf X}(t)\r$ with energy $E_n({\bf X}(t))$. Now if
the initial state $|\psi(0)\r$ belongs to the $n$th eigenspace,
then after the circuit $C$ the final state $|\psi(T)\r$ is given
by
\begin{equation}\label{}
  |\psi(T)\r = \exp(i\gamma_n(C)) \exp\left( -
  \frac{i}{\hbar}\int_0^T E_n({\bf X}(t))dt\right) |\psi(0)\r\ ,
\end{equation}
where the Berry phase reads:
\begin{equation}\label{}
  \gamma_n(C) = - \int\!\int_{\partial\Sigma=C} F_n^{\rm q}({\bf X})\ ,
\end{equation}
and $F_n^{\rm q}$  (Berry's curvature) denotes the following
two-form on $\cal M$:
\begin{equation}\label{}
  F_n^{\rm q}({\bf X}) = {\rm Im}\, \la d_{\bf X}\, n;{\bf X}| \wedge|d_{\bf X}\,
  n;{\bf X}\r \ .
\end{equation}
Using semiclassical analysis Berry shown \cite{Berry-clas} that
\begin{equation}\label{q-c}
  \Delta\theta(I;C) = -\hbar\frac{\partial \gamma_n(C)}{\partial I} =
   -\frac{\partial \gamma_n(C)}{\partial n} \ ,
\end{equation}
where $n$ is considered as a continuous variable according to
Bohr--Sommerfeld quantization rule $I = \hbar(n + \mu)$, with
$\mu$ being the Maslov index.

Both two-forms $F^{\rm c}(I;{\bf X})$ and $F_n^{\rm q}({\bf X})$
live in the parameter space $\cal M$. Clearly, they are defined by
very different objects: `classical form' $F^{\rm c}(I;{\bf X})$
uses phase-space quantities $q(I,\theta;{\bf X})$ and
$p(I,\theta;{\bf X})$ whereas `quantum form' $F_n^{\rm q}({\bf
X})$ is defined in terms of Hilbert space eigenvectors $|n;{\bf
X}\r$. There is, however, equivalent formulation of quantum
mechanics which uses objects defined on the classical phase space
only \cite{Wigner}. There is a direct relation --- well known
Wigner--Weyl correspondence --- between functions $F=F(q,p)$ on
the classical phase space and self-adjoint operators $\hat{F}$ in
the system Hilbert space. If $|\psi\r$ is the state vector,  then
\begin{equation}\label{FF}
  \la \psi|\hat{F}|\psi \r = \int W_\psi(q,p)F(q,p)\, dqdp\ ,
\end{equation}
where $W_\psi=W_\psi(q,p)$ is a Wigner function corresponding to
$|\psi\r$. Moreover, this formulation is perfectly suited to
semiclassical analysis. It is well known  that Wigner function
corresponding to the eigenstate $|n\r$ of the Hamiltonian
$\hat{H}$ depends only on $I$ and not on $\theta$, i.e. $W_n(q,p)
= W_n(I)$. In the classical limit, i.e. $\hbar \rightarrow 0, n
\rightarrow\infty$ such that $n\hbar$ is constant and equals
$I_0$, the Wigner function $W_n$ is concentrated on the torus
$I_0$
\begin{equation}\label{}
  W_n(I) \ \longrightarrow\ \frac{1}{2\pi}\, \delta(I-I_0)\ .
\end{equation}
If the quantum system depends upon external parameters $\bf X$
which evolve adiabatically then $W_n$ is adiabatically constant,
or, using the language of the classical adiabatic theorem, $W_n$
defines an adiabatic invariant. Now, since Berry's curvature $F_n$
is a measurable quantity it may be expressed according to
\begin{equation}\label{}
  F_n^{\rm q} = \int W_n(I)\, \mbox{`classical\ quantity'}\, dI \
  .
\end{equation}
Clearly, this `classical quantity' has to be related with the
`classical two-form' $F^{\rm c}$. Moreover, in the classical limit
$F_n^{\rm q}$ and its classical counterpart $F^{\rm c}$ have to be
related according to (\ref{q-c}), that is,
\begin{equation}\label{}
 F_n^{\rm q}({\bf X}) = - \frac 1\hbar\, F^{\rm c}(I;{\bf X})\ .
\end{equation}
 The natural choice is therefore
\begin{equation}\label{BASIC}
 F_n^{\rm q}({\bf X}) = -2\pi \int W_n(I)F^{\rm c}(I;{\bf X}) \, dI \  .
\end{equation}

\noindent {\it Example.} As an example let us consider a
generalized harmonic oscillator \cite{Berry-clas} defined by
\begin{equation}\label{H-X}
  H(q,p;{\bf X}) = \frac 12 \left( Xq^2 + 2Yqp + Zp^2 \right) \ ,
\end{equation}
where the parameters ${\bf X}=(X,Y,Z) \in \mathbb{R}^3$ satisfy
$XZ>Y^2$ (this condition implies that the above system describes
oscillatory motion round elliptic contours in the two-dimensional
phase space $ \mathbb{R}^2$). One  shows \cite{Berry-clas}
\begin{equation}\label{}
  F^{\rm c}(I;{\bf X}) = -
  \frac{I}{4\omega^3}\, (XdY
  \wedge dZ + YdZ \wedge dX + ZdX  \wedge dY )\ ,
\end{equation}
with the frequency of the quasi-periodic motion $  \omega =
\sqrt{XZ - Y^2}$. The quantized system (according to the
Wigner--Weyl correspondence) is given by
\begin{equation}\label{}
\hat{H}({\bf X}) = \frac 12 \left( X\hat{q}^2 + Y(\hat{q}\hat{p} +
\hat{p}\hat{q}) + Z\hat{p}^2 \right) \ .
\end{equation}
The eigen-equation $\hat{H}\psi_n=E_n\psi_n$ is solved by the
following normalized eigenfunctions:
\begin{equation*}\label{}
  \psi_n(q;{\bf X}) =  \sqrt{\alpha}
  \, \chi_n\left( \alpha q  \right)\,
  \exp\left(\frac{-iYq^2}{2Z\hbar} \right) \ ,
\end{equation*}
where $\alpha = \sqrt{{\omega}/{Z\hbar}}$, and
 $ \chi_n(\xi) = N_n e^{-\xi^2/2} H_n(\xi)$,
with $H_n$ being the $n$th Hermite polynomial and the
normalization constant $N_n = (2^n n! \sqrt{\pi})^{-1/2}$. Energy
levels are given by the standard formula  $E_n = \hbar\omega (n +
1/2)$.  The corresponding Wigner function $W_n(q,p;{\bf X})$ reads
as follows
\begin{eqnarray}\label{}
\lefteqn{  W_n(q,p;{\bf X}) = \frac{1}{\pi\hbar}
\int_{-\infty}^\infty\,
  ds\,
  \overline{\psi}_n(q+s;{\bf X}) {\psi}_n(q-s;{\bf X})\,
  e^{2ips/\hbar} \nonumber}  \\  &=&  \frac{\alpha}{\pi\hbar}
  \int_{-\infty}^\infty\, ds\,
  \overline{\chi}_n \left( \alpha (q+s)\right)
  {\chi}_n\left(  \alpha (q-s)\right)\,
  e^{2i[(p + Yq/Z)s]/\hbar} \nonumber \\ &=& W^{\rm
  osc}_n(q,\tilde{p})\ ,
\end{eqnarray}
where $W^{\rm osc}_n$ is the $n$th  Wigner function corresponding
to the standard harmonic oscillator
\begin{equation}\label{H-osc}
  H^{\rm osc}(q,\tilde{p}) = \frac 12 \left(  Z\tilde{p}^2 +
  \frac{\omega^2}{Z} q^2 \right) \ ,
\end{equation}
with $\tilde{p} = p + Yq/Z$. Clearly, the {\bf X}--dependent
canonical transformation $(q,p) \longrightarrow (q,\tilde{p})$
transforms (\ref{H-X}) into (\ref{H-osc}). Now, $W^{\rm osc}$ is
given by the well known formula (see e.g. \cite{Wigner})
\begin{equation}\label{}
W_n(I) =  W_n^{\rm osc}(q,\tilde{p}) =  \frac{(-1)^n}{\pi\hbar} \,
e^{-2I/\hbar}  L_n(4I/\hbar) \ ,
\end{equation}
where $I= H^{\rm osc}/\omega$ is the action variable and $L_n$
denotes the $n$th Laguerre polynomial. Finally, using
\begin{equation*}\label{}
  \int_0^\infty W_n(I)I\, dI = \frac{n + \frac 12}{2\pi} \ ,
\end{equation*}
one finds the following formula for the Berry curvature
\begin{eqnarray}\label{FB-oscillator}
  F_n^{\rm q}({\bf X}) &=& -   \frac{n+ \frac
  12}{I}\, F^{\rm c}(I;{\bf X}) \nonumber \\ &=&
  \frac{n+ \frac 12}{4\omega^3}\,  (XdY
  \wedge dZ + YdZ \wedge dX + ZX  \wedge dY )\ ,
\end{eqnarray}
in perfect agreement with \cite{Berry-clas}.

Our basic formula (\ref{BASIC}) may be generalized in two evident
ways: if the classical integrable system has $N$ degrees of
freedom then the corresponding Berry curvature reads:
\begin{equation}\label{BASIC-n}
 F_n^{\rm q}({\bf X}) = -(2\pi)^N \int\!\ldots\!\int
 W_n({\bf I})F^{\rm c}({\bf I};{\bf X}) \, d{\bf I} \ ,
\end{equation}
with
\begin{equation}\label{}
F^{\rm c}({\bf I};{\bf X}) =  \sum_{k=1}^N\la d_{\bf X}\, p_k({\bf
I} ;{\bf X}) \wedge d_{\bf X}\, q_k({\bf I};{\bf X})\r \ ,
\end{equation}
where now one averages over $N$-dimensional torus ${\bf I}=
(I_1,\ldots,I_N)$. The second generalization corresponds to the
non-abelian case developed by Wilczek and Zee \cite{WZ}. Suppose
that $n$th eigenvalue is $N$ times degenerate and let $|n,a;{\bf
X}\r$ $(a=1,\ldots,N$) span the corresponding $N$-dimensional
eigenspace. Then the Wilczek--Zee curvature is given by the
following formula
\begin{equation}\label{BASIC-WZ}
 F_{n;ab}^{\rm WZ}({\bf X}) = -2\pi \int W_{n;ab}(I)F^{\rm c}(I;{\bf X}) \, dI \
 ,
\end{equation}
with $W_{n;ab}$ being the following `Wigner matrix'
\begin{equation}\label{}
  W_{n;ab}(q,p;{\bf X})  = \frac{1}{\pi\hbar}
\int_{-\infty}^\infty\,
  ds\,
  \la n,a;{\bf X}|q+s\r \la q-s|n,b;{\bf X}\r \,
  e^{2ips/\hbar}\ .
\end{equation}
Clearly, $W_{n;ab}$ is hermitian and  hence $iF^{\rm WZ}_n\in
u(N)$. Now, changing $|n,a;{\bf X}\r$ to $| \widetilde{n,a};{\bf
X}\r = \sum_{b}U_{ab}({\bf X}) |n,b;{\bf X}\r$, with $U({\bf
X})\in U(N)$, one finds
\begin{equation}\label{}
  \widetilde{F}^{\rm WZ}_n({\bf X}) = U({\bf X}) F^{\rm WZ}_n({\bf X}) U^\dag({\bf
  X})\ ,
\end{equation}
that is, tensorial rule for the gauge transformation of $F^{\rm
WZ}_n$. Finally, the formula (\ref{BASIC}) suggests the following
generalization for the adiabatic evolution of mixed states.
Suppose that $\rho$ is a density operator such that the
corresponding Wigner function $W_\rho=W_\rho(I)$ is adiabatically
constant. Following Sj\"oqvist et. al. \cite{Sjoqvist} one defines
a phase of $\rho(T)$ with respect to $\rho$ as $\phi= {\rm arg}\,
{\rm Tr}[U(t)\rho]$. Now, $\phi$ may be recovered from space-phase
quantities as follows:
\begin{equation}\label{}
  \phi = \int\!\int_{\partial\Sigma=C} F_\rho({\bf X})\ ,
\end{equation}
where the two-form $F_\rho({\bf X})$ on the parameter space $\cal
M$ is defined by
\begin{equation}\label{rho-BASIC}
  F_\rho({\bf X}) = - 2\pi \int W_\rho(I) F^{\rm c}(I;{\bf X})\,
  dI\ .
\end{equation}
Clearly, if the stationary (in the adiabatic limit) state $\rho$
is pure, then necessarily $\rho = |n;{\bf X}\r\la n;{\bf X}|$ and
(\ref{rho-BASIC}) reduces to (\ref{BASIC}).

\vspace{.5cm}

\noindent {Acknowledgments:} This work was partially supported by
the Polish Ministry of Scientific Research and Information
Technology under the  grant No PBZ-MIN-008/P03/2003.

\end{document}